\documentclass{article}
\usepackage{
amssymb}
\usepackage{cite}                

\usepackage[latin9]{inputenc}
\input epsf          

\def\al{\alpha}

\def\la{\lambda}

\def\IC{\relax{\rm l\kern-.50 em C}}
\def\IE{\relax{\rm l\kern-.12 em E}}
\def\IK{\relax{\rm l\kern-.18 em K}}
\def\IL{\relax{\rm I\kern-.18 em L}}
\def\IN{\relax{\rm I\kern-.18 em N}}
\def\IR{\relax{\rm I\kern-.18 em R}}

\font\tenfrak=eufm10  \font\sevenfrak=eufm7  \font\fivefrak=eufm5
\newfam\frakfam
\textfont\frakfam=\tenfrak
\scriptfont\frakfam=\sevenfrak
\scriptscriptfont\frakfam=\fivefrak

\def\frac#1#2{{#1\over #2}}
\def\dfrac#1#2{{\displaystyle{#1\over #2}}}
\def\fracpd#1#2{\frac{\partial #1}{\partial #2}}

\def\ket#1{|#1\rangle}
\begin{document}
\title{ One-dimensional model of a Quantum  \\nonlinear Harmonic
 Oscillator\footnote{Presented to the 36 Symposium
 on Mathematical Physics, Toru\ \'n 9--12 June 2004 }}
\author{
 {\sc J.  F. Cari\~nena}\\ 
{\small
Departamento de F\'{\i}sica Te\'orica, Facultad de Ciencias,}\\
{\small
Universidad de Zaragoza,
50009 Zaragoza, Spain}\\ {\small
e-mail: jfc@posta.unizar.es }\\[2ex]
{\sc
Manuel F. Ra\~nada}\\ {\small
Departamento de F\'{\i}sica Te\'orica, Facultad de Ciencias,} \\ {\small
Universidad de Zaragoza,
50009 Zaragoza, Spain} \\ {\small
e-mail: mfran@unizar.es }
\\and\\ {\sc 
Mariano Santander}\\ {\small
 Departamento de F\'{\i}sica Te\'orica, Facultad de Ciencias} \\ {\small
  Universidad de Valladolid,  47011 Valladolid, Spain}\\{\small
e-mail: santander@fta.uva.es}
\\
}
\maketitle
\date{}
\begin{abstract}
In this paper we study the quantization of the nonlinear oscillator
introduced by Mathews and Lakshmanan. This system with
position-dependent mass allows a natural quantization procedure and
is shown to display shape invariance. Its energy spectrum is found by
factorization. The linear harmonic oscillator appears as the
$\lambda\to0$ limit of this nonlinear oscillator, whose energy
spectrum and eigenfunctions are  compared to the linear ones.

\medskip


\medskip

\end{abstract}

\section{Introduction}
In a recent paper \cite{CRSS04}  we have analyzed 
a classical nonlinear oscillator which is a  
$n$-dimensional generalization   of the one-dimensional system introduced
 previously  by 
  Mathews and Lakshmanan  \cite{MaL74},\cite{LaRa03},
 as a one-dimensional analogue of 
some models of quantum field theory \cite{DeSS69},\cite{NiW72}.
Such 
mechanical system was described by a Lagrangian
\begin{equation}
  L  = \frac{1}{2}\,\Bigl(\frac{1}{1 + \la\,x^2} \Bigr)\,(\dot{x}^2 - 
\al^2\,x^2)\ ,
\label{Lagn1}\end{equation}
which  represents a nonlinear oscillator-like
 with  amplitude dependent frequency
periodic solutions.
Note  that this system can also  be considered as an oscillator
with a position-dependent effective mass $m=(1 + \la\,x^2)^{-1} $
(see e.g. \cite{Le95} and \cite{KK03} and references therein).

The two-dimensional generalization studied in  \cite{CRSS04} was given by the Lagrangian 
\begin{equation}
  L(\lambda) = \frac{1}{2}\,\left(\frac{1}{1 + \la\,r^2}\right)\,
  \Bigl[\,v_x^2 + v_y^2 + \la\,(x\, v_y - y\, v_x)^2-\alpha^2\, r^2 \,\Bigr] \,,\quad
  r^2 = x^2+y^2     \,,
\label{Ln2}\end{equation}
and it was shown to be not only integrable but super-integrable. This suggests
 that  the corresponding quantum model should be exactly solvable, although one may expect to have some ordering ambiguities because of the $x$-dependence of the kinetic term. 

  We aim to study in this paper the one-dimensional quantum model using the 
well-known techniques of factorization and related operators (see e.g. \cite{CMPR98} and \cite{CR00} and references therein). This algebraic technique
was started by Schr\"odinger  in \cite{Sch1} and \cite{Sch2} and its
 interest has been increasing since the beginning of
Supersymmetric Quantum Mechanics (see \cite{CKS95} for a review). 

In more detail, the plan of the article is as follows:
Section 2 is devoted to study the simplest $\alpha=0$ case
in the classical approach \cite{CRSS04}, both in the Lagrangian and the Hamiltonian formalism,  and an  infinitesimal symmetry as well as
 the invariant measure in $\mathbb{R}$ under such vector field are
 determined.
Then, we proceed to introduce the quantum Hamiltonian describing this
position-dependent free system. The factorization method for this 
kind of position-dependent mass \cite{IH} is developed in Section 3
and the specific example of the quantum nonlinear oscillator
is studied in Section 4, where we prove that the problem has shape 
invariance. In Section 5 the spectrum of the quantum nonlinear oscillator is 
found by using the method proposed for such systems 
by  Gendenshte\"{\i}n in
\cite{Ge83,GK85}. The last section includes some final comments 
on the relation of this problem with that of the harmonic oscillator 
in the limit $\lambda\to 0$.

\section{$\lambda$-dependent ``Free Particle"}

  Let us first recall the case  of the one-dimensional ``free-particle''
motion (in the sense that $\alpha=0$) characterized by the Lagrangian
\begin{equation}
   L(x,v_x,\lambda) = T_1(\lambda)  = \frac{1}{2}\,\Bigl(\frac{v_x^2}{1 + 
\la\,x^2} \Bigr)\,.
\end{equation}
As it was remarked in \cite{CRSS04}, the  Lagrangian function $T_1(\lambda)$ is invariant under the action of the vector field
$X_x=X_x(\lambda)$ given by
$$
  X_x(\lambda) = \sqrt{\,1+\la\,x^2\,}\,\,\fracpd{}{x}  \,,
$$
in the sense that we have
$$
  X_x^t(\lambda)\Bigl(T_1(\lambda)\Bigr)=0  \,,
$$
where $X_x^t(\lambda)$ denotes the natural lift to the phase space $\mathbb{R}{\times}\mathbb{R}$
(tangent bundle in differential geometric terms) of the vector field 
$X_x(\lambda)$,
$$
  X_x^t(\lambda) = \sqrt{\,1+\la\,x^2\,}\,\,\fracpd{}{x}
  +  \Bigl(\frac{\la\,x\, v_x}{\sqrt{1+\la\,x^2\,}}\Bigr)\fracpd{}{v_x} \,.
$$

This vector field can be seen as a Killing vector field for the metric 
$g=(1 + \la\,x^2)^{-1}\, dx\otimes dx$, and generates a one-parameter group of isometries
 in this Riemann space. The natural measure in the real line is not 
invariant under such vector field; instead,  the only invariant measures are
 the multiples of 
$d\mu=(1 + \la\,x^2)^{-1/2}\, dx$. 

It is important to remark that in the space $L^2(\mathbb{R},d\mu)$ 
of square integrable functions in the real line,  the adjoint of 
the differential operator $\sqrt{\,1+\la\,x^2\,}\,\partial/\partial{x}$ is 
precisely 
the opposite of such operator.

On the other side, with the momentum defined as usual, 
$$p=\frac{\partial L}{\partial v_x}=\frac {v_x}{1 + \la\,x^2}\ ,
$$
the Legendre
transformation
$(x,v_x)\mapsto (x,p)$ leads to a Hamiltonian function given by 
$$H=(1 + \la\,x^2)\, \frac{p^2}2=\frac 12 \left(\sqrt{1 + \la\,x^2}\,p
\right)^2\ .
$$
 
Contrarily to the naive expectation of ordering ambiguities, however, 
the usual procedure of canonical quantization does not present
 any ambiguity because, as pointed out before, the linear operator
 (we put $\hbar=1$)
$$P=-i\,  \sqrt{1 + \la\,x^2}\,\fracpd{}x
$$
is self adjoint in the space  $L^2(\mathbb{R},d\mu)$,
and therefore the quantum Hamiltonian operator turns out to be
$$\widehat H=\frac 12\,P^2=-\frac 12 \left( \sqrt{1 + \la\,x^2}\,\fracpd{}x\right)^2
=-\frac 12\,(1 + \la\,x^2)\,\fracpd{^2}{x^2}-\frac 12\lambda\,x\,\fracpd{} x\ .
$$

In presence of an interaction described by a potential $V_1(x)$ things
works similarly and the Hamiltonian is then
$$ 
\widehat{H}_1 = -\frac 12\,(1 + \la\,x^2)\,\fracpd{^2}{x^2}
     -\frac 12\,\la\,x\,\fracpd{}{x} +V_1(x)  \,.
$$

\section{The factorization method}

Hereafter, as the configuration space is one-dimensional,  we will use 
the notation $d/dx$ instead of the more traditional for the vector field 
$\partial/\partial x$.

Let us try to determine a function $W(x)$,  called super-potential function, 
in such a way that the operator $A$ and its adjoint operator $A^{\dag}$,
given by 
\begin{eqnarray}
  A  &=  \dfrac 1{\sqrt 2}\left(\sqrt{\,1 + \la\,x^2\,}\ \dfrac{d}{dx} +W(x) 
\right)  \,,\cr
A^{\dag} &=\dfrac 1{\sqrt 2}\left( -\sqrt{\,1 + \la\,x^2\,}\ \dfrac{d}{dx} +W(x)\right)   \,,\nonumber
\end{eqnarray} 
are such that $\widehat H_1=A^{\dag} \,A$ \cite{IH}, i.e.
$$
  \widehat H_1 = A^{\dag}\,A =\frac 12
\Bigl[\,-\,\sqrt{\,1 + \la\,x^2\,}\,\frac{d}{dx} +W(x)\, \Bigr]
\Bigl[\,\sqrt{\,1 + \la\,x^2\,}\,\frac{d}{dx} +W(x)\, \Bigr]
$$

 Therefore, the  super-potential function $W$
must satisfy the following Riccati type differential equation 
$$
   \sqrt{\,1 + \la\,x^2\,}\ W' - W^2 +2\, V_1 = 0\ .
$$

Once such a factorization is obtained, we can define a new quantum 
Hamiltonian operator
$$ \widehat H_2 = A\,A^{\dag} =\frac 12 \,\Bigl[\,\sqrt{\,1 + \la\,x^2\,}\,\frac{d}{dx} +W(x)\, \Bigr]\,\Bigl[\,-\,\sqrt{\,1 + \la\,x^2\,}\,\frac{d}{dx} +W(x)\, \Bigr]\ ,
$$
which is called the partner Hamiltonian. The new potential $V_2$ is given in 
terms of the super-potential $W$ by 
$$
V_2=\frac 12\left(\sqrt{\,1 + \la\,x^2\,}\,W' + W^2\right) \ .
$$

The operator $A$ is such that 
$A\, \widehat H_1 = \widehat H_2 \,A$ while $A^{\dag}$ is such that 
$A^{\dag}\, \widehat H_2= \widehat H_1\,A^{\dag} $. This shows that if 
$\ket\Psi$ is an eigenvector of $\widehat H_1$ corresponding to the eigenvalue
$E$, then  when  $A\ket\Psi\neq 0$,  $A\ket\Psi$ is an eigenvector of  $\widehat H_2$ corresponding to the same eigenvalue, because 
$$
\widehat H_2\,A\,\ket\Psi=A\, \widehat H_1\,\ket\Psi=E\,A\,\ket\Psi\ ,
$$
and similarly, when $\ket\Phi$ is an eigenvector of $\widehat H_2$ corresponding to the eigenvalue $E$ and such that $A^{\dag}\,\ket\Phi\neq 0$, then 
$A^{\dag}\,\ket\Phi$ is an eigenvector of  $\widehat H_1$ corresponding to the 
same eigenvalue $E$. In other words, the spectra of $\widehat H_1$ 
and $\widehat H_2$ are almost identical, the only difference appearing when 
either $\ket\Psi$ is an eigenvector of $\widehat H_1$ but  
$A\,\ket\Psi= 0$,
or $\ket\Phi$  is an eigenvector of  $\widehat H_2$ for which 
$A^{\dag}\,\ket\Phi= 0$.

As an important first remark, the potential function $V$ is only defined 
up to addition of a constant, and therefore all the preceding expressions
 can be extended to the case in which $H$ is replaced by $H+c$, where $c$ is
 a constant. As another remark, it happens very often that some parameters may appear in the expression of the potential function, and therefore the  
super-potential function $W$ will also depend of the values
 of such parameters. 
The most important case is when the explicit forms of 
the potential and its partner are quite similar and only differ in
 the values of the
parameters, and then the problem is said to have shape invariance. 
As we will see this is the case for the quantum nonlinear oscillator 
we are considering in this paper.

\section{The quantum nonlinear  Oscillator}

In the particular case of the nonlinear harmonic oscillator for which 
the Hamiltonian is given by 
$$
H  = \frac{1}{2 }\,\left[(1 + \la\,x^2)\,p_x^2
     +\frac{\alpha^2\,x^2}{1 + \la\,x^2}\right] \ ,
$$
the quantum Hamiltonian operator will be
\begin{equation}
\widehat H_1  = \frac{1}{2 }\,\left[-(1 + \la\,x^2)\,
\frac{d^2}{dx^2}-\lambda\, x\, \frac d{dx}+\frac{\alpha^2\,x^2}{1 +
\la\,x^2}\right]\ .\label{eqH1}
\end{equation}
Now, if for any real number $\beta$ we define the linear operator in $L^2({\mathbb{R}},d\mu)$ 
\begin{equation}
  A  =  \dfrac 1{\sqrt 2}\left(\sqrt{\,1 + \la\,x^2\,}\ \dfrac{d}{dx} +\dfrac{\beta\,x}{\sqrt{1 + \la\,x^2}} 
\right)  \,,
\end{equation}
for which its adjoint operator is
\begin{equation}
A^{\dag} =\dfrac 1{\sqrt 2}\left( -\sqrt{\,1 + \la\,x^2\,}\ \dfrac{d}{dx}  +\dfrac{\beta\,x}{\sqrt{1 + \la\,x^2}} \right)   \,,
\end{equation}
then, we find  that 
$$A^{\dag} A = -\,\frac{1}{2}\,(1 + \la\,x^2)\,\frac{d^2}{dx^2}
     -\,\frac{1}{2}\,\la\,x\,\frac{d}{dx}
     +\,\frac{1}{2}\,\beta\,(\beta+\lambda)\,\left(\frac{x^2}{1 + 
\la\,x^2}\right)
     -\,\frac{1}{2}\,\beta  $$
where by simple comparing with (\ref{eqH1}) we conclude that the hamiltonian   
$\widehat H'_1=\widehat H_1-(1/2)\beta$ admits a factorization 
\begin{equation}
\widehat H'_1:=A^{\dag}\,A \label{eqH1fact}
\end{equation}
whenever
the parameters $\alpha$ and $\beta $ are related by 
$\alpha^2=\beta(\beta+\lambda)$, i.e.
$$\beta=-\frac 12\left(\lambda- \sqrt{\lambda^2+ 4 \alpha^2}\right)\ .
$$
Note that 
$$\lim_{\lambda\to 0}\beta=\alpha\ .
$$
Now for the partner hamiltonian  $\widehat H'_2:=A\,A^{\dag}$ we find 
\begin{equation}
\widehat H'_2=A\,A^{\dag}= -\,\frac{1}{2}\,(1 + \la\,x^2)\,\frac{d^2}{dx^2}
     -\,\frac{1}{2}\,\la\,x\,\frac{d}{dx}
     +\,\frac{1}{2}\,\beta\,(\beta-\lambda)\,\left(\frac{x^2}{1 + 
\la\,x^2}\right)
     +\,\frac{1}{2}\,\beta   \ . \label{supar}
\end{equation}
 
Coming back to the general case, we recall that when a quantum Hamiltonian $\widehat H_1(\alpha)$ depending on a
 parameter $\alpha$ admits a factorization in such a way that the 
partner Hamiltonian $\widehat H_2(\alpha)$ is of the same form as 
$\widehat H_1(\alpha)$ but for a different value of the parameter 
$\alpha$, it is usually said that  there is shape invariance. More 
specifically,  the quantum Hamiltonian $\widehat H_1(\alpha)$ 
admitting a factorization $\widehat H_1(\alpha)=A^{\dag}(\alpha)\,A(\alpha)$
is said to be shape invariant when there exists a function $f$ such that
for the partner $\widehat H_2(\alpha)=A(\alpha)\,A^{\dag}(\alpha)$ we
have:
\begin{equation}
\widehat H_2(\alpha)=\widehat H_1(\alpha_1)+R(\alpha_1)\ ,
\label{recrel}
\end{equation}
where  $\alpha_1=f(\alpha)$ and  $R(\alpha)$ is a constant depending on the parameter   $\alpha$. In this case, it was shown in \cite{Ge83} and \cite{GK85} that there is a method for exactly computing all the spectrum of $\widehat H_1$
(see e.g. \cite{CR00} for a modern approach). The bound state
$\ket{\Psi_0(\alpha)}$ is found by solving 
 $A(\alpha)\ket{\Psi_0(\alpha)}=0$, and has a zero 
energy. Then, using 
 (\ref{recrel}) we can see that $\ket{\Psi_0(\alpha_1)}$ is an eigenstate 
of $\widehat H_2(\alpha)$ with an  energy $E_1=R(\alpha_1)$,
because
\begin{equation}
\widehat H_2(\alpha)\ket{\Psi_0(\alpha_1)}=(\widehat H_1(\alpha_1)+R(\alpha_1))
\ket{\Psi_0(\alpha_1)}=R(\alpha_1)\ket{\Psi_0(\alpha_1)}\,.
\end{equation} 

Then,  $A^{\dag}(\alpha)\ket{\Psi_0(\alpha_1)}$ is the first excited state
of  $\widehat H_1(\alpha)$, with an  energy $E_1=R(\alpha_1)$, because
\begin{eqnarray}
\widehat H_1(\alpha)A^{\dag}(\alpha)\ket{\Psi_0(\alpha_1)}&=&
A^{\dag}(\alpha)\widehat H_2(\alpha)\ket{\Psi_0(\alpha_1)}=
A^{\dag}(\alpha)(\widehat H_1(\alpha_1)+R(\alpha_1))\ket{\Psi_0(\alpha_1)}\cr
&=&
R(\alpha_1)A^{\dag}(\alpha)\ket{\Psi_0(\alpha_1)}\,.\nonumber
\end{eqnarray}

This process should be iterated and we will find the sequence of energies for 
 $\widehat H_1(\alpha) $
\begin{equation}
E_k= \sum_{j=1}^k R(\alpha_j),\qquad E_0=0\,,
\end{equation}
the corresponding eigenfunctions being 
\begin{equation}
\ket{\Psi_n(\alpha_0)}=A^{\dag}(\alpha_0)A^{\dag}(\alpha_1)\cdots
A^{\dag}(\alpha_{n-1})\ket{\Psi_0(\alpha_n)}\,,
\end{equation}
where $\alpha_0=\alpha$ and  $\alpha_{j+1}=f(\alpha_j)$, namely,
 $\alpha_k=f^k(\alpha_0)=f^k(\alpha)$.

Now we can apply this process to the case we were considering, the parameter being $\beta$, because 
when comparing $\widehat H'_1$ given by (\ref{eqH1fact}) with its partner 
$(\ref{supar})$ we see that as
$$ \widehat  H'_1(\beta-\lambda) = -\,\frac{1}{2}\,\Bigl[\,(1 + \la\,x^2)\,\frac{d^2}{dx^2}
    +\,\la\,x\,\frac{d}{dx} \,\Bigr]
    +\,\frac{1}{2}\,(\beta-\lambda)\,\beta\,\left(\frac{x^2}{1 + 
\la\,x^2}\right)
  -\,\frac{1}{2}\,(\beta-\lambda)\,,
$$
then 
$$
   \widehat  H'_1(\beta-\lambda) = \Bigl[ \widehat H'_2(\beta)  -\,\bigl(\frac{1}{2}\bigr)\,\beta\,\Bigr]
  -\,\frac{1}{2}\,(\beta-\lambda)   \ ,
$$
and, therefore,
$$ \widehat H'_2(\beta)  = \widehat  H'_1(f(\beta)) + \beta -\frac{1}{2}
\,\lambda   \ , 
$$
where $f$ is the function $f(\beta) = \beta-\la$. If $R$ is the function defined by $R(\beta) = \beta + (1/2)\,\lambda$, then we see that 
$$
\widehat  H'_2(\beta) = \widehat H'_1(\beta_1) + R(\beta_1)\,.
$$ 
Therefore, as the quantum nonlinear oscillator we are considering has a shape 
invariance, we can develop the method sketched before for finding both the spectrum and the corresponding eigenvectors. 

The usual choice of parameters for these shape invariant systems is such that the function $f$ corresponds to a displacement by one unit and then this
 suggests us to use the parameter $\gamma=\beta/\lambda$ instead of $\beta$.

\section{The spectrum of the quantum nonlinear oscillator}
Our starting point should be the bound state $\ket{\Psi_0(\beta)}$
of the Hamiltonian $\widehat H'_1$. This eigenvector is determined by 
the condition $A(\beta)\,\ket{\Psi_0(\beta)}  =  0 $. More specifically, 
we should solve the differential equation
$$\frac{d}{dx}\,\Psi_0 +\,\beta\,\left(\frac{x}{1 + 
\la\,x^2}\right)\,\Psi_0 =  0   
$$
and therefore the wave function of the fundamental state 
 must be proportional to 
$$
  \Psi_0 (x)=  \frac{1}{\,(\,1 + \la\,x^2\,)^{\,r_0}} \,,\qquad r_0 = \frac
\beta{2\lambda}
$$

The energies of the excited states will be 
$$E'_1=R(\beta_1)=\beta-\lambda+\lambda/2=\beta-\lambda/2\ ,
$$
and iterating the process we get
$$E'_n=\sum_{k=1}^nR(\beta_k)=\sum_{k=1}^n\left(\beta_k+\frac\lambda 2\right)
=\sum_{k=1}^n\left(\beta-\lambda\,k+\frac\lambda 2\right)\ ,$$
and therefore,
$$E'_n=n\,\beta+\lambda\left[\frac n2-\sum_{k=1}^n k\right]=n\,\beta-\frac {n^2}2\,\lambda\,.$$

The energy of the eigenstates of $\widehat H_1=\widehat H'_1+(1/2)\beta$
will be given by 
$$E_n=n\,\beta-\frac {n^2}2\,\lambda+\frac 12 \beta\,.$$

This relation can also be written as
$$E_n=-\frac \lambda 2\left(n-\frac \beta \lambda \right)^2+\frac{\beta\,(\beta-\lambda)}{2\,\lambda}\ .
$$

The method developed in \cite{Ge83} and \cite{GK85} also provides us
the corresponding eigenfunctions as 
$$\Psi_1=A^{\dag}(\beta)\Psi_0(\beta_1),\quad \ldots
\ldots\quad,\quad \Psi_n=A^{\dag}(\beta)A^{\dag}(\beta_1)\cdots A^{\dag}(\beta_{n-1})\Psi_0(\beta_n)\ .$$

There is a clear difference between the cases $\lambda>0$ and $\lambda<0$. In fact, 
let us first remark  that the lowest value for $E'_n$ is $E'_0=0$. Therefore,
if $\lambda>0$ only those values $n$ are allowed for which
$$
\beta-\lambda\, \frac n2\geq 0\Longrightarrow n\leq\frac {2\beta}\lambda\ ,.
$$
On the contrary, when  $\lambda<0$ all natural numbers are allowed 
for $n$. 
It should  also be remarked that as long as $\lambda\neq0$, i.e, for both signs of $\lambda$,
the eigenvalues are not equally spaced.
\section{Final Comments and Outlook}

In this paper we have analyzed a quantum version of a nonlinear quantum 
oscillator with an amplitude dependent angular frequency. In the limit when 
the parameter $\lambda$ tends to zero   we recover the usual harmonic
 oscillator. So, we first note that if $\lambda=0$, then $\alpha=\beta$,
and  if we take into account that
$$\lim_{\lambda\to 0}{(1+\lambda\,x^2)^{\beta/(2\lambda)}}
=\exp\left(\lim_{\lambda\to 0}\frac \beta{2\lambda}\,(1+\lambda\,x^2)\right)=e^{\frac 12 \alpha x^2}\ ,
$$
we see that the ground state we have found coincides with that of the 
corresponding quantum linear harmonic oscillator. The limit, when $\lambda$
 goes to $0$, of operators $A$
 and $A^{\dag}$ and the eigenvalues of the Hamiltonian 
are the corresponding annihilation and creation operators  and the energy eigenvalues for the harmonic oscillator.

Another interesting issue concerns the behaviour of the position of the
energy levels when the parameter $\lambda$ changes. We remarked that
equispacing of the levels, a characteristic property of the usual
harmonic oscillator, no longer holds when $\lambda\neq0$. In
particular, the fundamental level is given by $E_0=\beta/2$; while a
superficial reading would suggest this is a constant independent of
$\lambda$ we must stress this is not the case, as $\beta$ itself has been
determined as a solution of the quadratic equation
$\alpha^2=\beta(\beta+\lambda)$, whose appropiate determination is 
$$
\beta=\frac{-\lambda+\sqrt{\lambda^2+4\alpha^2}}{2}
$$ 
Thus, if we consider the nonlinear
oscillator with fixed 'strenght' $\alpha$ and allow the
nonlinear parameter to vary, the fundamental level will actually depend
on $\lambda$.
The fundamental level $E_0$ (for $n=0$) has the standard oscillator value  $\alpha/2$
for $\lambda=0$, and is a positive and always decreasing function of $\lambda$. For large
negative values of $\lambda$ it approaches the large positive asymptotic regime
$E_0(\lambda) \approx -\lambda/2$, while for large positive values of $\lambda$ it tends to
zero.
  
It is a well-known fact that canonical transformation and  quantization do 
not commute.  We have made a canonical point transformation in such a way that 
the new coordinate is the one rectifying the vector field generating the isometries of the metric associated to the Lagrangian describing the system and then
the corresponding invariant measure is but the differential of such variable
and the Hamiltonian becomes the square of the momentum, eliminating then 
order ambiguities. With this quantization procedure we have found that the spectrum and the corresponding eigenvectors can be easily found using the fact
that the Hamiltonian admits such factorization that it is shape-invariant.
 
{\small
\section*{\bf Acknowledgments.}
Support of projects  BFM-2003-02532, FPA-2003-02948,
BFM-2002-03773, and CO2-399 is acknowledged.
 }
\end{document}